\documentclass[a4paper,10pt]{article}
\pdfoutput=1
\usepackage[top=1.0in, bottom=1.0in, left=1.0in, right=1.0in]{geometry}
\usepackage{amsmath}
\usepackage{graphicx}
\usepackage{helvet}
\usepackage{caption}
\usepackage{enumerate}

\def\feii{[Fe~{\textsc{ii}}]}

\def\nii{[N~{\textsc{ii}}]}
\def\halpha{H$\alpha$}

\def\brg{Br$\gamma$}
\def\fetwoline{[Fe~{\textsc{ii}}]\, 1.644\,$\mu$m}
\def\fetwoothers{[Fe~{\textsc{ii}}]\, 1.534+}
\def\ntwoline{[N~{\textsc{ii}}]\,$\lambda6583$}

\def\xfesun{X_\odot({\rm Fe/H})}
\def\fluxbrg{F_{{\rm Br}\gamma}}
\def\fluxfetwo{F_{\rm [Fe~{\textsc{ii}}]\,1.644}}
\def\fluxfetwoothers{F_{\rm [Fe~{\textsc{ii}}]\,1.534+}}
\def\simlt{\lower.5ex\hbox{$\; \buildrel < \over \sim \;$}}
\def\simgt{\lower.5ex\hbox{$\; \buildrel > \over \sim \;$}}
\def\arcdeg{\mbox{$^\circ$}}
\def\farcs{\mbox{$.\!\!^{\prime\prime}$}}
\def\fsecs{\mbox{$.\!\!^{\rm s}$}}
\def\kms{km s$^{-1}$}
\def\mum{$\mu$m}

\def\msun{{$M_\odot$}}
\def\vlsr{v_{\rm LSR}}
\def\vfwhm{\Delta v_{\rm FWHM}}
\def\sigmath{\sigma_{\rm th}}
\def\sigmanth{\sigma_{\rm nth}}
\def\sigmaobs{\sigma_{\rm obs}}

\def\nhcol{N_{\rm H}}
\def\fluxnarrow{F(1.644)_{\rm narrow}}
\def\fluxbroad{F(1.644)_{\rm broad}}
\def\fluxbrgnarrow{F({\rm Br}\gamma)_{\rm narrow}}

\begin{document}

\noindent\textbf{\textsf{\Large 
Detection of Pristine Circumstellar Material of the Cassiopeia A Supernova
}}

\vspace{7mm}
\noindent
Bon-Chul Koo$^{1\ast}$, Hyun-Jeong Kim$^{1,2}$, Heeyoung Oh$^{3,4}$, John C. Raymond$^{5}$, Sung-Chul Yoon$^{1}$, Yong-Hyun Lee$^{1,3}$, Daniel T. Jaffe$^{4}$

\vspace{7mm}
\noindent
$^{1}$Department of Physics and Astronomy, Seoul National University, Seoul 08826, Korea\\
$^{2}$Department of Astronomy and Space Science, Kyung Hee University, 
Yongin-si, Gyeonggi-do 17104, Korea\\
$^{3}$Korea Astronomy and Space Science Institute, Daejeon 34055, Korea\\
$^{4}$Department of Astronomy, University of Texas at Austin, Austin, TX 78712, USA\\
$^{5}$Harvard-Smithsonian Center for Astrophysics, 60 Garden Street, Cambridge, MA 02138, USA

\vspace{7mm}
{\bf \noindent
Cassiopeia A is a nearby young supernova remnant that provides 
a unique laboratory for the study of core-collapse supernova explosions$^{1}$. 
Cassiopeia A is known to be a Type IIb supernova from the optical 
spectrum of its light echo$^{2}$, but 
the immediate progenitor of the supernova remains uncertain$^{3}$. 
Here we report results of near-infrared, high-resolution spectroscopic  
observations of Cassiopeia A  where we detected the pristine circumstellar material 
of the supernova progenitor. Our observations revealed a strong 
emission line of iron (Fe) from a circumstellar clump that has not yet been processed by 
the supernova shock wave. 
A comprehensive analysis of the observed spectra, together with 
an {\em HST} image, indicates that the 
majority of Fe in this unprocessed circumstellar material is in the gas phase, 
not depleted onto dust grains as in the general interstellar medium$^{4}$. 
This result is consistent with a theoretical model$^{5,6}$ of dust condensation 
in material that is heavily enriched with CNO-cycle products, supporting the idea 
that the clump originated near the He core of the progenitor$^{7,8}$. 
It has been recently found that Type IIb supernovae can result from 
the explosion of a blue supergiant with a thin hydrogen envelope$^{9-11}$, 
and our results support such a  scenario for Cassiopeia A. 
}
\bigskip

Core-collapse supernovae and their young remnants interact with 
the circumstellar material (CSM) ejected at the end of the progenitors' lifetime$^{12}$. 
By studying the physical and chemical characteristics of this material, 
we can learn how the progenitors stripped off their envelopes and exploded, 
which is crucial for understanding the nature of progenitors.

Cassiopeia A (Cas A) is a young ($\sim 340$ yr)$^{13}$ supernova remnant (SNR) 
where we observe the interaction of  the SNR blast wave with the CSM. 
Its SN type is Type IIb, indicating that the progenitor had a 
thin H envelope at the time of explosion$^{2}$. 
The morphology and expansion rates of the Cas A SNR suggest that 
it is interacting with a smooth red supergiant (RSG) wind$^{14,15}$. 
The X-ray characteristics of the shocked ejecta knots and shocked ambient gas 
are also consistent with Cas A expanding into an RSG wind$^{16,17}$. 
On the other hand, an X-ray spectral analysis indicates that there could have been 
a small bubble produced by a fast tenuous wind in the post-RSG stage$^{16}$. 
Hence, it is uncertain whether the Cas A SN exploded in 
an RSG phase with the dense slow wind extending all the way 
to the stellar surface or if, instead, there was a short 
blue phase with a faster wind just prior to the explosion$^{3}$. 

A distinct component of the CSM in Cas A is the so-called 
``quasi-stationary floculli (QSFs)'' (Fig. 1). 
These nebulosities or clumps are almost `stationary'  ($\simlt 400$~\kms) 
and are bright in H$\alpha$ and \nii$\, \lambda\lambda$6548, 6583 
emission line images$^{18-22}$. 
Their optical and near-infrared (NIR) spectra indicate that 
QSFs are dense (3--$9 \times 10^4~\rm cm^{-3}$) and 
He and N enriched$^{7,23-25}$. 
QSFs are probably dense CNO-processed circumstellar clumps that 
have been shocked recently by the SN blast wave$^{14,26}$. 
It was pointed out that a progenitor of 15--25 \msun\ must 
have lost $\sim 60\%$ of its mass prior to the ejection of 
such N-rich QSFs$^{8}$, but the evolutionary stage of the progenitor 
and the mass loss mechanism are uncertain. 
It had been suggested that QSFs are the fragments of an RSG shell 
formed by a fast wind in the progenitor's Wolf-Rayet phase$^{27}$, but hydrodynamic 
simulations showed that QSFs are not consistent with such fragments$^{15}$. 
Another suggestion was that QSFs are the clumps of an inhomogeneous RSG wind$^{14}$, 
but it is not clear how the He- and N-rich clumps ejected from 
the bottom of H envelope can be at their current locations.

We report results of high-resolution, NIR spectroscopic observations 
of one of QSFs in Cas A where we detected the emission 
from unshocked, pristine CSM. 
The emission from shock-processed CSM in QSFs has been observed 
since the 1950s$^{7,18,21-25}$, but the emission from the 
unprocessed CSM that retains the original physical and chemical conditions of 
the deep interior of the progenitor has never been observed. 
There have been observations of patchy optical emission outside the SNR 
that could be also an unprocessed CSM, but 
from the outer envelope of the progenitor ejected during the RSG phase$^{14,28}$. 
The knot (hereafter `Knot 24'$^{29}$) 
that we observed lies near the  southern radio boundary of the SNR, 
at the tip of a prominent arc of QSFs (Fig. 1), and 
has been visible at least since 1951$^{20,21}$. 
The observations were performed with 
the Immersion GRating INfrared Spectrometer (IGRINS) 
mounted on the 4.3m Discovery Channel Telescope (DCT) at Lowell Observatory 
in 2018 December.

We detected a strong \fetwoline\ line with a remarkable profile showing 
a very narrow line superposed on a very broad one (Fig. 2a). 
The broad-line component (hereafter `BLC') has a velocity width of 
$\sim 200$~\kms, indicating that it is emitted from shocked gas. 
The width of the narrow-line component (hereafter `NLC') 
is $\sim 8$~\kms. 
(See Methods and Supplementary Table 1 for the line parameters.) 
The entire knot revealed in the NLC map extends $6''$ NW-SE 
and has a complex morphology with several clumps 
including `Clump A-NLC', the largest one in the southeast (Fig. 3a). 
The northwestern boundary of Clump A-NLC touches 
and lies immediately to the southeast of 
a bright emission feature in the BLC map labelled `Clump A-BLC' (Fig. 3b). 
The detailed structure of Clump A-BLC can be 
seen in the {\it HST} F625W image in Fig. 3c, which is 
dominated by the H$\alpha$ and \nii$\,\lambda\lambda$6548, 6583 
lines from radiative shock. 
The prominent emission feature in the middle of the {\it HST} image is 
spatially coincident with Clump A-BLC and shows a sharp decrease of brightness 
towards Clump A-NLC. This morphology suggests that 
a strong shock (hereafter the `NW shock') is currently propagating into 
Clump A from NW to SE and that Clump A-NLC is the unshocked part of the clump. 
We can also see faint \halpha+\nii\ emission along 
the NE rim of Clump A-NLC, implying another, weaker, shock propagating 
into the clump from NE to SW. 
There are other emission features in the NLC and BLC maps 
to the northwest of Clump A, but the \feii\ emission is faint and 
the relation between the narrow- and broad-line components is unclear. 

We detected other \feii\ lines at 1.534, 1.600, and 1.677 \mum\ 
and the hydrogen \brg\ line toward Clump A (Fig. 2). 
The electron density of Clump A-NLC inferred from the ratio of \feii\ lines is 
$1,000 \pm 700$~cm$^{-3}$, which is more than an order of magnitude lower than 
that of the BLC ($1.8\pm 0.2 \times 10^4$~cm$^{-3}$; Supplementary Table 1). 
Another way to estimate the density of Clump A-NLC is to consider 
pressure equilibrium. Since the typical temperature of 
the \feii-line emitting post-shock cooling layer is $T_e=7,000$ K and 
the ionization fraction is $\sim 0.5$ (ref.$^{30}$), 
the pressure of the shocked gas becomes 
$\approx 3 n_e T_e\approx 3.8\times 10^8$ cm$^{-3}$ K. 
The ram pressure of the preshock gas should be comparable to this; 
the preshock density therefore would be 
$n_{c}=220 (v_s/100~{\rm km~s}^{-1})^{-2}$ cm$^{-3}$. 
The temperature of Clump A-NLC inferred from the line widths of 
the \brg\ ($\sim 23$~\kms) and \fetwoline\ ($\sim 8.5$~\kms) 
lines is $T=(1.2\pm 0.5)\times 10^4$~K (Methods). 
An order of magnitude lower density in the NLC than in 
the shocked gas and the estimated temperature of $\sim10^4$~K, together 
with the morphology in Fig. 3, indicate that 
Clump A-NLC is unshocked, {\it pristine} CSM photoionized by UV radiation 
from the NW shock propagating into the clump.

The narrow \fetwoline\ line detected toward Clump A-NLC is strong, 
with the ratio of \brg\ to \fetwoline\ fluxes ($\fluxbrg/\fluxfetwo$) of 
$0.10\pm 0.03$. This ratio is much smaller than that seen toward nebulae 
where Fe is believed to be heavily depleted on 
dust grains (e.g., 7--10 in the Orion bar$^{31}$), suggesting that 
Fe in Clump A-NLC is mostly in the gas phase. 
We have analyzed the observed line ratios toward Clump A using a shock model (Methods). 
The analysis shows that the observed ratio of the NLC to BLC \fetwoline\ line 
flux ($\approx 0.12$) is consistent with a shock speed of 100--125~\kms, 
which agrees with the shock speed implied by the line width of 
the BLC ($\simgt 100$~\kms) as well as the shock speed of the NW shock front 
obtained by comparing the 2018 {\it HST} image to another {\it HST} image 
taken in 2004 March (100--150~\kms; Supplementary Fig. 1). 
While the calculated $\fluxbrg/\fluxfetwo$ ratio that assumed 
 solar abundance agrees with the observed ratio for the BLC, 
it is $0.6(\pm 0.2)$ times the observed ratio for the NLC. 
The simplest interpretation, therefore, is that, in the unshocked CSM, 
the Fe abundance relative to H by number is 60$(\pm 20)$\% of solar, while, 
in the shocked gas, it is equal to solar, presumably because Fe locked up in 
dust grains is released by the shock destruction of the grains. 
A CSM with the majority of Fe in the gas phase is 
in sharp contrast to the general interstellar medium 
where $\le 5$\% of Fe is in the gas phase$^{4}$. 

The `non-depletion' of Fe in Knot 24 must reflect the physical and 
chemical conditions of the stellar material ejected 
from the progenitor in the pre-SN stage. 
The N and He overabundance in QSFs indicates that the QSFs originated 
from a N-rich layer at the bottom of the H envelope of an RSG (Fig. 4)$^{7,8}$. 
Knot 24 is as N enriched as other QSFs; its \ntwoline /\halpha\ ($3.0\pm 0.2$) is  
close to the mean value for QSFs (3.3; ref.$^{22}$). 
Hence, Knot 24 might have originated with the other  
QSFs from the same N-rich layer. 
Note that QSFs are He overabundant relative to H by a factor of 
4--10 (ref.$^{7}$), suggesting that they originated near the He core 
where the material is heavily enriched with CNO products as 
in the surface of WNL stars$^{7,8}$. 
In that thin layer, the abundances of C and O drop considerably below 
their solar values, while the abundances of heavier refractory 
elements (e.g., Mg, Si, and Fe) remain essentially the same. 
The composition of dust formed from such CNO-processed material 
would be different from 
that of dust formed in the outer envelope of RSG winds because 
neither the oxygen rich minerals of M stars nor the mixture of carbon 
dust and carbides as in C stars can be formed$^{6}$.  
Theoretical model calculations found that the composition of dust formed in such material 
is dominated by solid Fe and FeSi alloys and that, 
as a consequence of the low probability of Fe grain formation, 
as much as 80--85\% of Fe is left in the gas phase$^{5,6,32}$. 
Hence, the non-depletion of Fe in Knot 24 is consistent with 
dust condensation in a CSM heavily enriched with CNO products. 

When did the mass loss that ejected the QSFs occur 
and what was the evolutionary stage of the progenitor at that time? 
Proper motion studies of QSFs in Cas A found a systemic ``expanding'' motion 
with a characteristic time of ($1.1\pm0.2) \times 10^4$~yr and 
an expansion velocity of 180~\kms\ (when scaled to 3.4 kpc)$^{19,20}$. 
If the apparent expanding motion is partly due to shock motion ($\sim 100$~\kms), 
the systemic expansion velocity could have been overestimated by 
a factor of $\sim 2$. 
The corresponding radius and the surface temperature of the progenitor 
should be about 100~$R_\odot$ and $\sim 10^4$ K, respectively, 
if the progenitor mass at that stage were about 5~\msun\ (Methods). 
These values, together with the He and N overabundance of QSFs, imply that 
the immediate progenitor of Cas A SN was likely a blue supergiant (BSG) with 
a WNL-like surface chemical composition rather than an RSG. 
If the expansion velocity of QSFs is lower than 100~\kms, 
a yellow supergiant could be also possible. 
The optical spectrum of Cas A SN  implies that the H envelope 
mass in the Cas A SN progenitor was comparable to that of 
SN 1993J (i.e., $M_\mathrm{H,env} \approx 0.1~M_\odot$; ref.$^{2}$). 
Although the SN 1993J progenitor was identified as an RSG$^{33}$, 
recent theoretical and observational studies indicate that both RSG and BSG solutions 
for Type IIb SN progenitors can be found with $M_\mathrm{H,env} \sim 0.1~M_\odot$, 
for which the surface abundances are similar to WNL stars$^{9-11}$. 
Therefore, the scenario in which the He- and N-rich, Fe-non-depleted, 
QSFs originated from a BSG wind emitted from the progenitor is consistent 
with recent theoretical and observational findings. 
The formation and dynamical evolution of such dense clumps in this scenario 
need to be explored. 

\vspace{7mm}
{\parindent0pt
{\bf\large References}\\
1. Milisavljevic, D. \& Fesen, R. A. The supernova - supernova remnant connection. In Alsabti, A. W. \& Murdin, P. (eds.) {\it Handbook of Supernovae}, 2211-2231 (Springer International Publishing AG, 2017).\\
2. Krause, O. {\it et al.} The Cassiopeia A supernova was of Type IIb. {\it Science} {\bf 320,} 1195-1197 (2008).\\
3. Chevalier, R. A. \& Soderberg, A. M. Type IIb supernovae with compact and extended progenitors. {\it Astrophys. J. Lett.} {\bf 711,} L40-L43 (2010).\\
4. Savage, B. D. \& Sembach, K. R. Interstellar abundances from absorption-line observations with the Hubble space telescope. {\it Annu. Rev. Astron. Astr.} {\bf 34,} 279-330 (1996).\\
5. Gail, H. P., Duschl, W. J., Ferrarotti, A. S. \& Weis, K. Dust formation in LBV envelopes. In Humphreys, R. \& Stanek, K. (eds.) {\it The Fate of the Most Massive Stars, vol. 332 of Astronomical Society of the Pacific Conference Series}, 317-319 (Astronomical Society of the Pacific, San Francisco, 2005).\\
6. Gail, H. P. Formation and evolution of minerals in accretion disks and stellar outflows. In Henning, T. (ed.) {\it Astromineralogy, vol. 815}, 61-141 (Springer, Berlin, 2010).\\
7. Chevalier, R. A. \& Kirshner, R. P. Spectra of Cassiopeia A. II. Interpretation. {\it Astrophys. J.} {\bf 219,} 931-941 (1978).\\
8. Lamb, S. A. The Cassiopeia A progenitor: a consistent evolutionary picture involving supergiant mass loss. {\it Astrophys. J.} {\bf 220,} 186-192 (1978).\\
9. Meynet, G. {\it et al.} Impact of mass-loss on the evolution and pre-supernova properties of red supergiants. {\it Astron. Astrophys.} {\bf 575,} A60 (2015).\\
10. Yoon, S.-C., Dessart, L. \& Clocchiatti, A. Type Ib and IIb supernova progenitors in interacting binary systems. {\it Astrophys. J.} {\bf 840,} 10 (2017).\\
11. Kilpatrick, C. D. {\it  et al.} On the progenitor of the Type IIb supernova 2016gkg. {\it Mon. Not. R. Astron. Soc.} {\bf 465,} 4650-4657 (2017).\\
12. Smith, N. Mass loss: its effect on the evolution and fate of high-mass stars. {\it Annu. Rev. Astron. Astr.} {\bf 52,} 487-528 (2014).\\
13. Thorstensen, J. R., Fesen, R. A. \& van den Bergh, S. The expansion center and dynamical age of the galactic supernova remnant Cassiopeia A. {\it Astron. J.} {\bf 122,} 297-307 (2001).\\
14. Chevalier, R. A. \& Oishi, J. Cassiopeia A and its clumpy presupernova wind. {\it Astrophys. J. Lett.} {\bf 593,} L23-L26 (2003).\\
15. van Veelen, B., Langer, N., Vink, J., Garc{\'\i}a-Segura, G. \& van Marle, A. J. The hydrodynamics of the supernova remnant Cassiopeia A. The influence of the progenitor evolution on the velocity structure and clumping. {\it Astron. Astrophys.} {\bf 503,} 495-503 (2009).\\
16. Hwang, U. \& Laming, J. M. The circumstellar medium of Cassiopeia A inferred from the outer ejecta knot properties. {\it Astrophys. J.} {\bf 703,} 883-893 (2009).\\
17. Lee, J.-J., Park, S., Hughes, J. P. \& Slane, P. O. X-Ray observation of the shocked red supergiant wind of Cassiopeia A. {\it Astrophys. J.} {\bf 789,} 7 (2014).\\
18. Baade, W. \& Minkowski, R. Identification of the radio sources in Cassiopeia, Cygnus A, and Puppis A. {\it Astrophys. J.} {\bf 119,} 206-214 (1954).\\
19. Kamper, K. \& van den Bergh, S. Optical studies of Cassiopeia A. V. A definitive study of proper motions. {\it Astrophys. J. Suppl. S.} {\bf 32,} 351-366 (1976).\\
20. van den Bergh, S. \& Kamper, K. Optical studies of Cassiopeia A. VII. Recent observations of the structure and evolution of the nebulosity. {\it Astrophys. J.} {\bf 293,} 537-541 (1985).\\
21. Lawrence, S. S. {\it et al.} Three-dimensional Fabry-Perot imaging spectroscopy of the Crab nebula, Cassiopeia A, and nova GK Persei. {\it Astron. J.} {\bf 109,} 2635-2893 (1995).\\
22. Alarie, A., Bilodeau, A. \& Drissen, L. A hyperspectral view of Cassiopeia A. {\it Mon. Not. R. Astron. Soc.} {\bf 441,} 2996-3008 (2014).\\
23. Hurford, A. P. \& Fesen, R. A. Reddening measurements and physical conditions for Cassiopeia A from optical and near-infrared spectra. {\it Astrophys. J.} {\bf 469,} 246-254 (1996).\\
24. Gerardy, C. L. \& Fesen, R. A. Near-infrared spectroscopy of the Cassiopeia A and Kepler supernova remnants. {\it Astron. J.} {\bf 121,} 2781-2791 (2001).\\
25. Lee, Y.-H., Koo, B.-C., Moon, D.-S., Burton, M. G. \& Lee, J.-J. Near-infrared knots and dense Fe ejecta in the Cassiopeia A supernova remnant. {\it Astrophys. J.} {\bf 837,} 118 (2017).\\
26.  McKee, C. F. \& Cowie, L. L. The interaction between the blast wave of a supernova remnant and interstellar clouds. {\it Astrophys. J.} {\bf 195,} 715-725 (1975).\\
27. Chevalier, R. A. \& Liang, E. P. The interaction of supernovae with circumstellar bubbles. {\it Astrophys. J.} {\bf 344,} 332-340 (1989).\\
28. Fesen, R. A., Becker, R. H. \& Blair, W. P. Discovery of fast-moving nitrogen-rich ejecta in the supernova remnant Cassiopeia A. {\it Astrophys. J.} {\bf 313,} 378-388 (1987).\\
29. Koo, B.-C. {\it et al.} A deep near-infrared [Fe II]+[Si I] emission line image of the supernova remnant Cassiopeia A. {\it Astrophys. J.} {\bf 866,} 139 (2018).\\
30. Koo, B.-C., Raymond, J. C. \& Kim, H.-J. Infrared [Fe II] emission lines from radiative atomic shocks. {\it J. Kor. Astron. Soc.} {\bf 49,} 109-122 (2016).\\
31. Walmsley, C. M., Natta, A., Oliva, E. \& Testi, L. The structure of the Orion bar. {\it Astron. Astrophys.} {\bf 364,} 301-317 (2000).\\
32. Morris, P. W. {\it et al.} $\eta$ Carinae's dusty homunculus nebula from near-infrared to submillimeter wavelengths: mass, composition, and evidence for fading opacity. {\it Astrophys. J.} {\bf 842,} 79 (2017).\\
33. Aldering, G., Humphreys, R. M. \& Richmond, M. SN 1993J: the optical properties of its progenitor. {\it Astron. J.} {\bf 107,} 662-672 (1994).\\
34 Reed, J. E., Hester, J. J., Fabian, A. C. \& Winkler, P. F. The three-dimensional structure of the Cassiopeia A supernova remnant. I. The spherical shell. {\it Astrophys. J.} {\bf 440,} 706-721 (1995).\\
35. Paxton, B. {\it et al.} Modules for experiments in stellar astrophysics (MESA). {\it Astrophys. J. Suppl. S.} {\bf 192,} 3 (2011).\\

{\bf\large Correspondence}\\
Correspondence and requests for materials should be addressed to B.-C. K.~(email: koo@astro.snu.ac.kr). \\

{\bf\large Acknowledgements}\\
We wish to thank Rob Fesen for his helpful comments on an earlier version of the manuscript.  
This research was supported by Basic Science Research Program through the National Research Foundation of Korea(NRF) funded by the Ministry of Science, ICT and future Planning (2017R1A2A2A05001337).
This work used the Immersion Grating Infrared Spectrometer (IGRINS) that was developed under a collaboration between the University of Texas at Austin and the Korea Astronomy and Space Science Institute (KASI) with the financial support of the US National Science Foundation under grant AST-1229522, of the University of Texas at Austin, and of the Korean GMT Project of KASI.
These results made use of the Discovery Channel Telescope at Lowell Observatory. Lowell is a private, non-profit institution dedicated to astrophysical research and public appreciation of astronomy and operates the DCT in partnership with Boston University, the University of Maryland, the University of Toledo, Northern Arizona University and Yale University. \\

{\bf\large Author Contributions}\\
All authors contributed to different aspects of the project, and read and commented on the manuscript. 
B.-C.K. led the project, analysis and discussion, and wrote the manuscript. 
H.-J.K. performed the observation and data reduction, and contributed to the data analysis and manuscript writing. 
H.O. performed the observation and contributed to the IGRINS data analysis.
J.C.R. contributed to the shock emission analysis and scientific interpretation.
S.-C.Y. contributed to the scientific interpretation and the manuscript writing. 
Y.-H.L. contributed to the HST data analysis.  
D.T.J. contributed to the project setup. \\

{\bf\large Competing Interests}\\ 
The authors declare that they have no competing financial interests.
}

\clearpage

\begin{figure}
\includegraphics[width=1\textwidth]{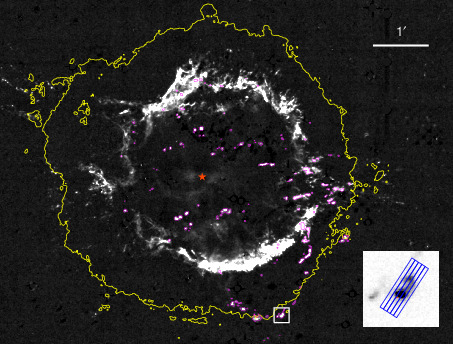}
\linespread{1.0}\selectfont{}
\caption{{\bf A deep [Fe II] 1.644~\mum\ image of Cas A$^{29}$.} 
North is up and east is to the left. QSFs are marked by magenta contours. 
The QSF marked by the white box in the southern area is Knot 24 
(= R37 of ref.$^{20}$ = QSF 3 of ref.$^{22}$). 
The inset shows a zoomed-in view of Knot 24 and the locations of 
the slits for our spectral mapping observation. 
The red star represents the explosion center at 
$(\alpha,\delta)_{\rm J2000}=(23^{\rm h} 23^{\rm m} 27\fsecs77, +58\arcdeg 48' 49\farcs4)$$^{13}$, 
while the yellow contour marks the outer boundary of the SNR in radio. 
The scale bar in the upper right represents an angular scale of $1'$, 
which corresponds to $\sim 1$ pc at the distance (3.4 kpc; ref.$^{22,34}$) of the SNR.
}
\end{figure}

\begin{figure}
\begin{center}
\includegraphics[width=0.55\textwidth]{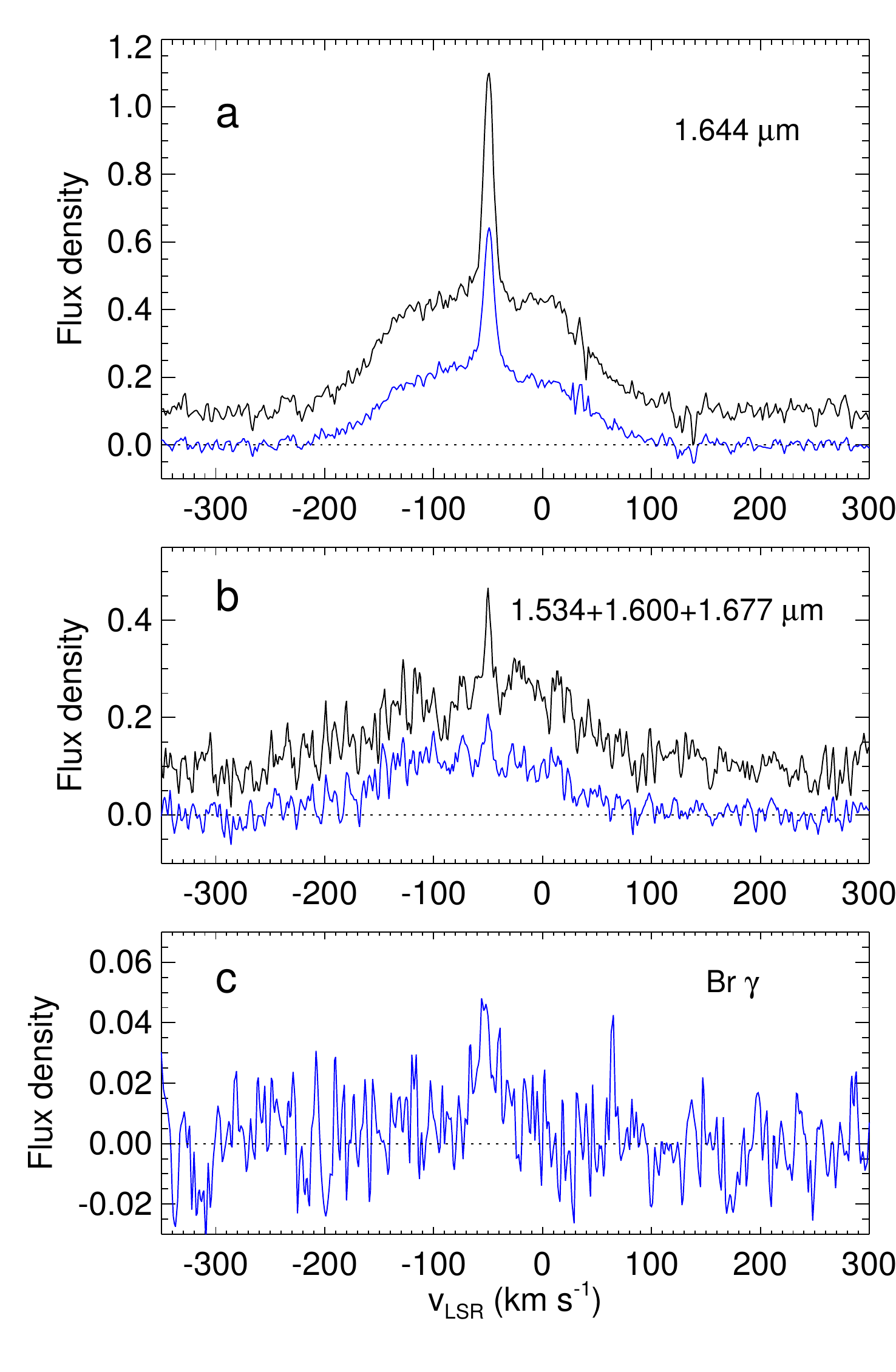} 
\linespread{1.0}\selectfont{}
\caption{{\bf Average spectra of Knot 24.} 
The spectra of the entire knot are shown in black (shifted upwards by 0.1) and 
those of Clump A in Fig. 3 are shown in blue. 
{\bf a}, \fetwoline\ line. There are a few artifacts from OH line 
removal, e.g., at $\vlsr\sim +40$~\kms\ and +130~\kms. 
{\bf b}, Sum of \feii\ 1.534, 1.600, and 1.677 \mum\ line spectra. 
{\bf c}, \brg\ line. The \brg\ line is very faint and can be seen only in Clump A. 
The flux density is normalized by the maximum flux density of the \fetwoline\ line. 
}
\end{center}
\end{figure}

\begin{figure}
\begin{center}
\includegraphics[width=1.0\textwidth]{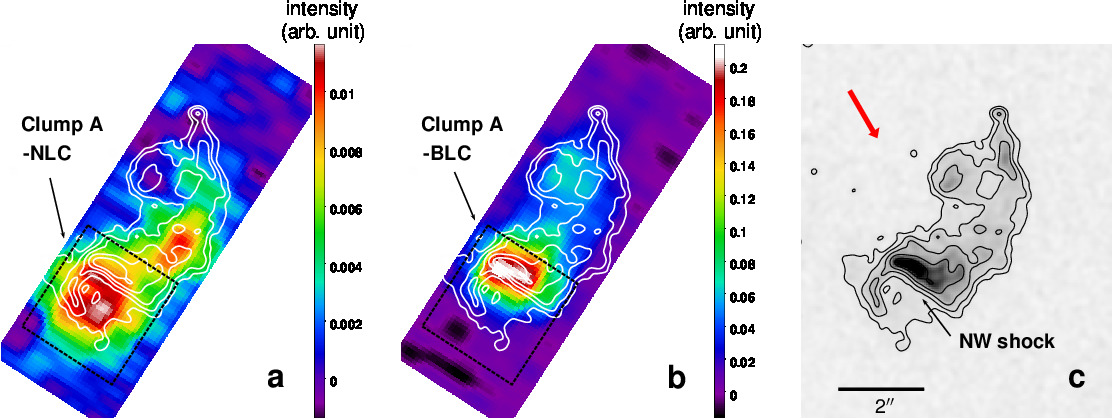} 
\linespread{1.0}\selectfont{}
\caption{{\bf Integrated intensity  [Fe II] 1.644~\mum\ maps of the narrow- and 
broad-line components of Knot 24 compared with an {\it HST} image.} 
{\bf a}, NLC map. The intensity scale is linear in arbitrary units. 
The dotted box marks the area used to derive the parameters of Clump A. 
{\bf b}, Same as {\bf a} but for the BLC. 
{\bf c}, {\it HST} WFC3/UVIS F625W image 
taken in 2018 January (Program ID 15337, P.I. Robert Fesen). 
The scale bar in the lower left corresponds to $2''$ (= 0.033 pc at 3.4 kpc). 
The thick red arrow shows the radial direction from the explosion center. 
The white contours overlaid in {\bf a} and {\bf b} are the contours of 
the {\it HST} image with intensity levels 0.25, 0.5, 1, 2, and 4 in arbitrary units. 
 }
\end{center}
\end{figure}

\begin{figure}
\begin{center}
\includegraphics[width=0.65\textwidth]{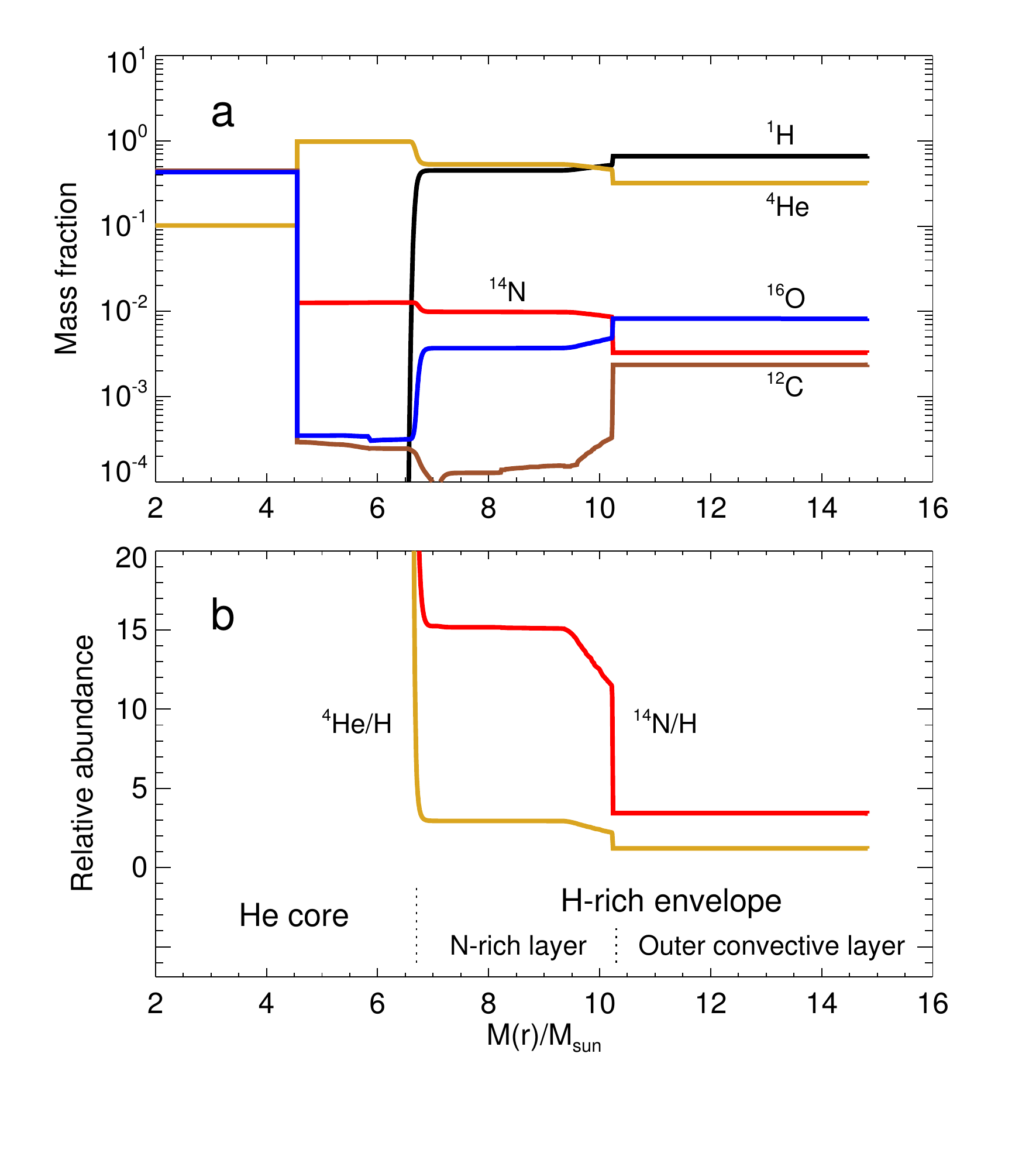} 
\linespread{1.0}\selectfont{}
\caption{{\bf Internal chemical structure of a non-rotating 20~\msun\ single star in the RSG stage.} 
The abscissa is the mass inside a radius $r$ from the explosion center. 
{\bf a}, Mass fraction of CNO cycle elements. 
{\bf b}, ${^4}$He/H and $^{14}$N/H ratio in mass. 
The ratios are normalized by their solar values. The initial metallicity is $Z = 0.02$ and 
the He mass fraction at the center of the He-burning core is  $Y_c=0.1$. 
The figure represents the structure about $2\times 10^5$~yr before the explosion, 
when the stripping of the outer layers of the H envelope would mostly occur. 
As He burning continues, the lower boundary of the outer convective layer in 
the H envelope moves further down, but the basic structure remains the same. 
The model presented here was  obtained by using the MESA code$^{35}$.
}
\end{center}
\end{figure}

\clearpage

\noindent{\bf\large Methods}\\

\noindent{\bf Observations and Data Reduction.} 
We obtained a high-resolution, NIR spectral map of Knot 24 using  
the Immersion GRating INfrared Spectrometer (IGRINS)$^{36-38}$ mounted on 
the 4.3m Discovery Channel Telescope (DCT) at Lowell Observatory 
on 2018 December 25 (UT). 
IGRINS simultaneously covers the {\it H} (1.47--1.81~\mum) and {\it K} (1.95--2.48~\mum) 
bands with a spectral resolving power of $R (\equiv \lambda/\Delta\lambda) \simeq 45,000$, 
corresponding to a velocity resolution of 7--8~\kms.  The slit size is $0\farcs63 \times 9\farcs3$. 
We observed five adjacent slit positions with an offset of $0\farcs63$ in 
the slit-width direction to make a spectral map of $3\farcs15 \times 9\farcs3$ covering 
the bright part of Knot 24 (Fig. 1). The position angle was 147$\arcdeg$, and 
the total on-source integration time varied from 15 min to 50 min depending on the slit position. 
The seeing in {\it K}-band measured from the IGRINS slit-view camera images 
was $0\farcs5$--$0\farcs7$. 
Between the on-source observations, sky frames were obtained at a position $6'$ east
of the on-source position for sky subtraction and wavelength calibration. 
We also observed the A0V-type star HR 9019 for telluric correction and flux calibration.

Data reduction was carried out with the IGRINS Pipeline Package (PLP)$^{39}$ v2.2.0, 
which performs sky subtraction, flat-fielding, bad-pixel correction, aperture extraction, 
and wavelength calibration. 
The PLP provides the 2D spectra of {\it H} and {\it K} bands with the 2D variance maps 
given by Poisson noise combined with the standard deviation of the frame produced by 
subtracting the off-source (sky) frames from the on-source frames. 
The wavelength solution is derived by fitting OH emission lines from 
the sky frames and has a typical uncertainty smaller than $\pm 0.5$~\kms. 
The PLP products were additionally processed by using the Plotspec 
python code designed for analyzing the reduced 2D IGRINS spectra$^{40}$. 
Plotspec performs telluric correction, relative flux calibration, and continuum subtraction 
and provides a single 2D spectrum with all orders in {\it H-} and {\it K}-bands combined. 
Using Plotspec, we obtained 2D position-velocity diagrams of the observed lines for each slit 
and a 3D datacube, which were used to generate average spectra and 
integrated intensity maps, respectively (Figs. 2 and 3).
In Fig. 3, the NLC map was produced by subtracting a baseline over the central 
velocity channels from the individual spectra, and the BLC map was obtained by 
subtracting the NLC map from the total integrated intensity map. \\

\noindent{\bf Derivation of Line Parameters.} 
We derived the \fetwoline\ line parameters of the NLC by fitting the observed spectra 
with a Gaussian profile and those of the BLC by a direct method after subtracting the narrow 
Gaussian component, i.e., the line flux from a direct sum, 
the central velocity $v_0$ from an intensity-weighted average, 
and the velocity width by dividing the line flux by the intensity at $v_0$. 
The parameters of Clump A were derived from the spectra 
extracted from the boxed area in Fig. 3. 
We find that Clump A contributes about 65\% (62\%) to 
the total narrow (broad) \fetwoline\ line flux of the entire knot. 
The derived line parameters are summarized in Supplementary Table 1. 

The other detected \feii\ lines at 1.534, 1.600, and 1.677 \mum\ have 
comparable critical densities of a few $10^4$~cm$^{-3}$, and 
their ratios to the \fetwoline\ line can be used as a density tracer. 
Since these lines are so weak that the NLC is barely visible in the individual spectra, 
we have added them to increase the signal-to-noise ratio. 
The resulting ``\fetwoothers'' spectrum of the entire knot is shown 
in Fig. 2b in which we can clearly see both NLC and BLC. 
The $\fluxfetwoothers/\fluxfetwo$ ratios are 0.17 and 0.46 for the NLC and BLC, 
respectively (Supplementary Table 1). For Clump A, the NLC 
of \fetwoothers\ line is relatively weak (Fig. 2b) 
with an $\fluxfetwoothers/\fluxfetwo$ ratio of 0.12. 
For the BLC, the ratio is 0.54. Supplementary Table 1 lists 
the electron densities derived from the $\fluxfetwoothers/\fluxfetwo$ ratios. 
For the BLC, it is (1.2--1.8)$\times 10^4$ cm$^{-3}$. 
For the NLC, the derived electron densities have large uncertainties, 
i.e., $\sim 2,000\pm 800$ cm$^{-3}$ (entire knot) and $\sim 1,000\pm 700$~cm$^{-3}$ (Clump A). 
The result is not sensitive to $T_e$ and we assumed $T_e=7,000$~K. 

The H \brg\ line is very faint and can be seen only in Clump A (Supplementary Fig. 2). 
In the average \brg\ spectrum of Clump A (Fig. 2c), the NLC is clearly seen 
with the line width significantly broader than the \fetwoline\ line (Fig. 2a). 
The BLC also appears to be present, but 
the baseline fluctuation hampers an accurate measurement of the line parameters. 
The parameters of the \brg\ line of Clump A were obtained by 
Gaussian fitting (for narrow line) and by the direct method (for broad line), 
and they are listed in Supplementary Table 2. 
To derive the ratio of the \brg\ line to \fetwoline\ line fluxes, we need to 
consider  the difference in extinction at two wavelengths. 
Estimates of the H column density to Knot 24 indicate that 
$N_{\rm H}=1.4\times 10^{22}$~cm$^{-2}$, corresponding  
to $A_V=7.4$ mag (ref.$^{29}$). 
If we use the dust opacity model of the general interstellar medium$^{41}$, 
$A_{1.64\,\mu{\rm m}}=1.37$~mag and $A_{2.17\,\mu{\rm m}}=0.85$~mag, so that 
the difference in the extinction at the two wavelengths is 0.5 mag, and 
we divided the observed ratio by 1.6. The resulting $\fluxbrg/\fluxfetwo$ ratios are 
$\sim 0.1$ and $\sim 0.02$ for the NLC and BLC, respectively 
(Supplementary Table 2). The latter, however, has a large uncertainty 
due to the baseline fluctuation. \\

\noindent{\bf Temperature and Non-thermal Motion of Clump A.} 
The line width (FWHM) of the narrow \brg\ line is $\sim 23$~\kms, 
much greater than that ($\sim 8.5$~\kms) of the narrow \fetwoline\ line 
(Supplementary Tables 1 and 2). 
These widths are corrected for the instrumental broadening ($\sim 8$~\kms). 
Note that the observed FWHM of the narrow \fetwoline\ line 
before the correction is 12~\kms, 
which is significantly greater than the instrumental resolution. 
Hence, the \fetwoline\ line is resolved, and its width is considerably greater than 
the thermal width ($\sim 3$~\kms) expected from the width of the \brg\ line, 
indicating that the line broadening is due to 
both thermal and temperature-independent non-thermal motions. 
We derived the temperature and non-thermal velocity dispersion of 
Clump A-NLC from the velocity widths of the narrow \brg\ and \fetwoline\ lines, 
assuming that the two lines originated in the regions 
of similar physical conditions.
The observed velocity dispersions ($\sigmaobs=\vfwhm/2.35$) 
are related to the thermal ($\sigmath$) and non-thermal ($\sigmanth$) velocity 
dispersions by 
\begin{equation}
\sigma_{\rm obs,\, Br\gamma}^2 = \sigmanth^2 + \sigma_{\rm th,\, Br\gamma}^2 \quad{\rm and} \quad
\sigma_{\rm obs,\, \feii}^2 = \sigmanth^2 + \sigma_{\rm th,\, \feii}^2, 
\end{equation}
where the thermal velocity dispersions of \brg\ and \fetwoline\ lines are given by
\begin{equation}
\sigma_{\rm th,\, Br\gamma}  = 9.12\, T_4^{1/2}~{\rm km~s}^{-1} \quad{\rm and} \quad
\sigma_{\rm th,\, \feii} =  \sigma_{\rm th,\, Br\gamma} /\sqrt{56} 
= 1.22\, T_4^{1/2}~~{\rm km~s}^{-1},  
\end{equation} 
respectively. In this equation, $T_4\equiv T/10^4~{\rm K}$. The observed velocity 
dispersions of the \brg\ and \fetwoline\ lines of Clump A-NLC are 
$\sigma_{\rm obs,\, Br\gamma}= 9.9\pm2.0$~\kms\ and 
$\sigma_{\rm obs,\, \feii}=3.6\pm0.2$~\kms, respectively 
(Supplementary Tables 1 and 2). Then, from equations (1) and (2), we obtained 
$\sigmanth=3.4\pm 0.2$~\kms\ and $T=(1.2\pm 0.5)\times 10^4$~K. \\

\noindent{\bf Shock Emission Analysis.} 
We compare the observed line flux ratios with the results from a shock model where 
the BLC is emitted from the shocked gas, while the NLC is emitted from the 
`radiative precursor', i.e., the region in front of the shock front heated/ionized by shock radiation. 
We calculated the \fetwoline\ and \brg\ line fluxes of the radiative precursor 
using the results in the MAPPINGS III Library of Fast Shock Models$^{42}$. 
The authors provide online tables summarizing the physical structure of 
the precursor and shock, i.e., temperature, density, and fraction of elements 
in different ionization stages as a function of distance from the shock front, 
for a range of shock parameters. 
We calculated the line emissivities using these tables 
with updated atomic parameters and integrated the emissivities 
to obtain line fluxes from the precursor region. 

For the atomic parameters of \feii\ forbidden lines, we used 
the effective collision strengths of ref.$^{43}$ and 
the radiative transition probabilities ($A$ values) of ref.$^{44}$. 
We note that the \fetwoline\ line fluxes in the MAPPINGS III Library 
can be reproduced by using the effective collision strengths and 
the $A$ values of refs.$^{45,46}$, 
and they are a factor of $\sim 3$ smaller than ours. 
The discrepancy mainly originates from the different collisional excitation rates.
The \feii\ atomic rates are still uncertain, but the main difference between 
the two sets of the collisional excitation rates is that ref.$^{43}$ includes 
the contribution of resonances to the excitation cross sections, and 
they dominate the total rates for most of the forbidden transitions. 
In ref.$^{30}$, we discussed recent updates. 
Note that the relative Fe abundance of two media, i.e., the unshocked and 
shocked CSM, depends only weakly on the adopted atomic parameters. 
Also, we adopted an Fe abundance of 
$\xfesun=3.47\times 10^{-5}$ in number$^{47}$, which is 
1.5 times greater than that adopted in the online table. 
Because of these differences in parameters, 
our \fetwoline\ line fluxes are a factor of $\sim 5$ 
greater than those in the MAPPINGS III Library. 
For the \brg\ line, we computed the flux of H$\beta$ line  
with the on-the-spot approximation and multiplied by 0.033 to derive 
the flux of \brg\ line, where the factor ``0.033'' is the ratio of the two line fluxes 
at $T=5,000$ K and $n_e\simlt 10^3$~cm$^{-3}$ (ref.$^{48}$). 
The results agree with those of the library to within a few percent. 

Supplementary Fig. 3a shows the physical structure of 
the precursor of a plane-parallel shock propagating at 125~\kms. 
The preshock density is $n_0=100$~cm$^{-3}$ and 
magnetic field strength is 0.1 $\mu$G. 
The abscissa is H-nucleus column density from the shock front ($\nhcol$). 
We see that the photoionized precursor 
extends to  $N({\rm H})\approx 2.6\times 10^{19}$ cm$^{-2}$. 
In this precursor region, the temperature is constant at 5,500~K 
and H is almost fully ionized. 
(The H ionization fraction closely follows the \brg\ line emissivity 
curve and is not shown.) 
The fraction of Fe in Fe$^+$ is 0.2 at the shock front 
and smoothly increases to 1 in the ambient medium. 
Near the shock front, Fe is mostly in Fe$^{++}$. 
Also shown are the normalized emissivities of \fetwoline\ and \brg\ lines. 
The two lines are both emitted from the entire precursor region, 
but the \fetwoline\ line emissivity attains a maximum further upstream 
because of the variation of the Fe$^+$ fraction. 
The structures of the precursors for other shock velocities are similar 
but with different spatial extents, e.g., $N({\rm H})=1.8\times 10^{19}$ cm$^{-2}$ 
and $7.5\times 10^{19}$~cm$^{-2}$ for 100~\kms\ and 150~\kms\ shocks. 
It is worthwhile to note that, in Clump A, the distribution of the \brg\ emission 
is slightly offset from that of Clump A-NLC in \fetwoline\ line 
(Supplementary Fig. 2), which appears consistent with this difference seen 
in the model. 

The column density of Clump A is 
$\nhcol\sim 2n_c R\sim 1\times 10^{19} n_{c,2}$ ~cm$^{-2}$ 
where $n_{c,2}=n_c/(100~{\rm cm}^{-3})$ and we used $R=0.016$~pc. 
If $n_c=200$~cm$^{-3}$, $\nhcol\sim2\times 10^{19}$~cm$^{-2}$ and it  
is less than the total column density of the radiative precursor of 
the 125~\kms\ shock in Supplementary Fig. 3a. 
In order to compare the result of the shock model to 
our observations, therefore, the precursor needs to be truncated. 
The \fetwoline\ line flux of such a truncated precursor can be obtained by 
integrating the emissivity from the shock front to a given $\nhcol$, i.e.,
\begin{equation}
F(1.644)_{\rm pre}(x_N) =2\pi \int_0^{x_N} j_{1.644}(x) \, dx
\end{equation}
where $j_{1.644}(x)$ is the volume emissivity of 
\fetwoline\ line at $x$, and $x_N$ is the distance corresponding to $\nhcol$. 
This is the flux normal to the shock front, but since we will be comparing 
the flux ratios, its absolute value is not relevant. 
Supplementary Fig.~3b shows the variation of the ratio 
$F(1.644)_{\rm pre}/F(1.644)_{\rm shock}$ for 100, 125, and 150~\kms\ shocks. 
The \fetwoline\ line flux from the shock ($F(1.644)_{\rm shock}$) is obtained by 
scaling the result in the MAPPINGS III library because we only need the total flux.
The observed flux ratio of the NLC to the BLC is 12\%, which agrees 
with the $F(1.644)_{\rm pre}/F(1.644)_{\rm shock}$ ratio of the 100--125~\kms\ shock 
for $N_{\rm H}=1$--$2\times 10^{19}$ ~cm$^{-2}$. 
The shock speed could be little higher depending on the density of Clump A. 

We also calculated the flux ratio of \brg\ to \fetwoline\ lines ($\fluxbrg/\fluxfetwo$) 
in the truncated precursor, and the result is shown in  
Supplementary Fig. 3c.
The observed $\fluxbrgnarrow/\fluxnarrow$ ratio ($0.10\pm 0.03$) is slightly 
larger than the ratio of the 100--125 \kms\ shock ($\sim 0.06$), 
but consistent within a factor of 2. 
For comparison, the observed $\fluxbrg/\fluxfetwo$ ratio 
of the BLC ($0.02\pm0.01$) agrees with the ratio of shock emission predicted 
from the model, e.g., 0.017 for the 125~\kms\ shock from the result 
in the MAPPINGS III library. 
In other words, the observed $\fluxbrg/\fluxfetwo$ ratio is comparable to 
the ratio expected for the solar abundance material in both  
shocked and unshocked CSM, implying that Fe is not significantly depleted 
in both media. According to the shock model presented here, 
the Fe abundance relative to H by number 
in the unshocked CSM is 60$(\pm 20)$\% of solar. 
There is some additional uncertainty in this fraction  
associated with atomic parameters. 
Note that if $\le 5$\% of Fe is in the gas phase as in the general interstellar 
medium$^{4}$, the $\fluxbrgnarrow/\fluxnarrow$ ratio of the precursor predicted 
from our shock model would be $\ge 1.2$, which is an order of magnitude 
greater than the observed $\fluxbrgnarrow/\fluxnarrow$ ratio. 
We also note that, since the {\em relative} $\fluxbrg/\fluxfetwo$ ratio in 
the two regions depends only weakly on the adopted \feii-line atomic parameters, 
the result that Fe in the unshocked CSM is not depleted more than 
a factor of $\sim 2$ than in the shocked CSM should be robust. \\

\noindent{\bf Estimate of the Radius and Surface Temperature of the Cas A Progenitor.} 
Given that the stellar wind velocity should be comparable to the escape velocity 
from the surface of the star (i.e., $v_\mathrm{w} \approx v_\mathrm{esc} = \sqrt{2GM_*/R_*}$), 
the progenitor radius can be given by 
\begin{equation} 
R_* \approx 50~R_\odot~
\left(\frac{v_\mathrm{w}}{200\, \mathrm{km~s^{-1}}}\right)^{-2}
\left(\frac{M_*}{5\, M_\odot}\right)~. 
\end{equation} 
The wind velocity of 100--200~$\mathrm{km~s^{-1}}$ inferred from 
the QSF proper motion implies $R_* \simeq 50$--$200\, R_\odot$, if the mass 
of the progenitor at the final evolutionary stage was about 5 $M_\odot$ (ref.$^{17}$). 
The stellar evolution models$^{10}$ predict that the luminosity of a Type IIb SN 
progenitor of $M_* \simeq 5 M_\odot$ would be about $L_* =10^5 L_\odot$ 
during the final evolution. From the blackbody approximation of 
$L_* = 4\pi R_*^2 \sigma T_*^4$, we get the surface temperature 
$T_* \simeq (0.7$--$1.4) \times 10^4~\mathrm{K}$. This temperature range 
corresponds to a BSG. If the wind velocity is lower than 100~\kms, the temperature 
could fall into the range of a yellow supergiant, e.g., 4,800--7,500~K (ref.$^{49}$). 
The inferred surface temperature depends only weakly on the assumed progenitor mass. 
For example, for $M_* = 3 M_\odot$, we get $R \simeq 30$--$120 R_\odot$, 
and using $L \simeq 10^{4.7}L_\odot$ (ref.$^{10}$), 
$T \simeq (0.8$--$1.6) \times 10^4~\mathrm{K}$. 

\vspace{7mm}
\noindent{\bf\large Data Availability}\\
The data that support the plots within this paper and other findings of this study are 
available from the corresponding author upon reasonable request.

\vspace{7mm}
{\parindent0pt
{\bf\large References only in Methods}\\
36. Yuk, I.-S. {\it et al.} Preliminary design of IGRINS (Immersion GRating INfrared Spectrograph). In McLean, I. S., Ramsay, S. K. \& Takami, H. (eds.) {\it Ground-based and Airborne Instrumentation for Astronomy III, vol. 7735 of Proceedings of SPIE}, 77351M (SPIE, Washington, 2010).\\
37. Park, C. {\it et al.} Design and early performance of IGRINS (Immersion Grating Infrared Spectrometer). In Ramsay, S. K., McLean, I. S. \& Takami, H. (eds.) {\it Ground-based and Airborne Instrumentation for Astronomy V, vol. 9147 of Proceedings of SPIE}, 91471D (SPIE, Washing- ton, 2014).\\
38. Mace, G. {\it et al.} IGRINS at the Discovery Channel Telescope and Gemini South. In Evans, C. J., Simard, L. \& Takami, H. (eds.) {\it Ground-based and Airborne Instrumentation for Astronomy VII, vol. 10702 of Proceedings of SPIE}, 107020Q (SPIE, Washington, 2018).\\
39. Lee, J.-J., Gullikson, K. \& Kaplan, K. F. IGRINS pipeline package (IGRINS/PLP) v2.2.0. \\
http://doi.org/10.5281/zenodo.845059 (2017).\\
40. Kaplan, K. F. {\it et al.} Excitation of molecular hydrogen in the Orion bar photodissociation region from a deep near-infrared IGRINS spectrum. {\it Astrophys. J.} {\bf 838,} 152 (2017).\\
41. Draine, B. T. Scattering by interstellar dust grains. I. Optical and ultraviolet. {\it Astrophys. J.} {\bf 598,} 1017-1025 (2003).\\
42. Allen, M. G., Groves, B. A., Dopita, M. A., Sutherland, R. S. \& Kewley, L. J. The MAPPINGS III library of fast radiative shock models. {\it Astrophys. J. Suppl. S.} {\bf 178,} 20-55 (2008).\\
43. Ramsbottom, C. A., Hudson, C. E., Norrington, P. H. \& Scott, M. P. Electron-impact excitation of Fe II. Collision strengths and effective collision strengths for low-lying fine-structure forbidden transitions. {\it Astron. Astrophys.} {\bf 475,} 765-769 (2007).\\
44. Deb, N. C. \& Hibbert, A. Radiative transition rates for the forbidden lines in Fe II. {\it Astron. Astrophys.} {\bf 536,} A74 (2011).\\
45. Nussbaumer, H. \& Storey, P. J. Atomic data for Fe II. {\it Astron. Astrophys.} {\bf 89,} 308-313 (1980).\\
46. Nussbaumer, H. \& Storey, P. J. Transition probabilities for Fe II infrared lines. {\it Astron. Astrophys.} {\bf 193,} 327-333 (1988).\\
47. Asplund, M., Grevesse, N., Sauval, A. J. \& Scott, P. The chemical composition of the Sun. {\it Annu. Rev. Astron. Astr.} {\bf 47,} 481-522 (2009).\\
48. Draine, B. T. {\it Physics of the Interstellar and Intergalactic Medium} (Princeton University Press, New Jersey, 2011).\\
49. Drout, M. R., Massey, P., Meynet, G., Tokarz, S. \& Caldwell, N. Yellow supergiants in the Andromeda galaxy (M31). {\it Astrophys. J.} {\bf 703,} 441-460 (2009). \\
}

\clearpage
\newpage
\noindent\textbf{\textsf{\large SUPPLEMENTARY INFORMATION}}

\vspace{10mm}

\noindent
This section contains all the supplementary data (2 tables and 3 figures) supporting the analysis presented in the main paper and the Method section.

\bigskip\bigskip

%\clearpage

%%%---Tables
\captionsetup[table]{name=Supplementary Table}

\begin{table}[h]
\centering
\linespread{1.0}\selectfont{}
\caption{\bf Parameters of [Fe~II] emission of Knot 24}
\begin{tabular}{cccc}
\hline
\hline
{ } & { } & {Entire knot} & {Clump A} \\
\hline
narrow line  &  $v_0$ (\kms)      & $-$49.68(0.31)   & $-$49.28(0.31)   \\
                    & $\vfwhm$ (\kms)  & 7.45(0.44)          & 8.45(0.47)          \\
                    & $\fluxfetwo$  & 10.92(0.25)          & 7.15(0.20)         \\
                    & $\fluxfetwoothers$ & 1.86(0.38)     & 0.87(0.29)         \\
       & $\fluxfetwoothers/\fluxfetwo$ & 0.171(0.035)  & 0.122(0.040)    \\
       &  $n_e$~($10^3$ cm$^{-3}$)  & 1.97($-$0.70, $+$0.80) & 1.05($-$0.64, $+$0.75)   \\
\\
broad line   &  $v_0$ (\kms)          & $-$56.1   &  $-$57.5  \\
                   & $\vfwhm$ (\kms)     & 205     & 189    \\
                   & $\fluxfetwo$       & 100               & 62.3(0.7)       \\
                   & $\fluxfetwoothers$  & 46.2(2.0)      & 33.9(1.5)        \\
       & $\fluxfetwoothers/\fluxfetwo$ & 0.462(0.021) & 0.543(0.025)  \\
       &  $n_e$~($10^4$ cm$^{-3}$)  & 1.23($-$0.12, $+$0.13)   & 1.81($-$0.20, $+$0.22)     \\
\\
\multicolumn{2}{c}{narrow to broad line $\fluxfetwo$  ratio} & 0.109(0.003) & 0.115(0.003)  \\
\hline 

\multicolumn{4}{p{0.9\textwidth}}{\rm{\textsc{Note}---$v_0, \vfwhm$ = velocity center and velocity width of \fetwoline\ line; 
$\fluxfetwo$ =\feii\ 1.644 \mum\ line flux;
$\fluxfetwoothers$ = sum of \feii\ 1.534, 1.600, and 1.677 \mum\ line fluxes; 
$n_e$ = electron density derived from $\fluxfetwoothers/\fluxfetwo$ 
assuming $T_e=7,000$~K.
The line fluxes are normalized to the \fetwoline\ flux (= 100) of the broad line of the entire knot.
For the narrow line, the line parameters are derived from a 
Gaussian fit and $\vfwhm$ is the FWHM corrected for the instrumental broadening ($8.2\pm0.4$~\kms).
The errors are $1\sigma$ errors.
For the broad line,  $v_0$ is the intensity weighted mean velocity
and $\vfwhm$ is the velocity width obtained by dividing the  
line flux by the intensity at $v_0$.
}}
\end{tabular}
\end{table}

%\clearpage
%\newpage
\bigskip\bigskip

\begin{table}[h]
\centering
\linespread{1.0}\selectfont{}
\caption{\bf Parameters of \brg\ emission of Clump A in Knot 24}
\begin{tabular}{ccc}
\hline
\hline
{ } & {parameter} & {value} \\
\hline
narrow line  &  $v_0$ (\kms)      & $-$51.4(1.9)  \\
                    & $\vfwhm$ (\kms)  & 23.2(4.8)        \\
                    & $\fluxbrg$  & 1.19(0.30)        \\
                    & $\fluxbrg/\fluxfetwo$ & 0.104(0.026)\textsuperscript{$\ast$} \\ 
\\
broad line   & $\fluxbrg$       & 1.89(0.52)              \\
                   & $\fluxbrg/\fluxfetwo$ & 0.020(0.010)\textsuperscript{$\ast$} \\ 
\hline
\multicolumn{3}{p{0.7\textwidth}}{\textsuperscript{$\ast$}{\rm These are extinction-corrected flux ratios (see Methods).}} \\
\multicolumn{3}{p{0.7\textwidth}}{\rm {\textsc{Note}---$v_0, \vfwhm$ = velocity center and velocity width; 
$\fluxbrg$ = \brg\ line flux normalized to the \fetwoline\ flux (= 100) of the broad line of the entire 
knot as in Supplementary Table 1.
The velocity width is corrected for the instrumental broadening ($8.2\pm0.9$~\kms).
The errors are $1\sigma$ errors.
}}
\end{tabular}
\end{table}

\clearpage
\newpage

%%%---Figures
\captionsetup[figure]{name=Supplementary Figure, skip=15pt}
\setcounter{figure}{0}

\begin{figure}
\begin{center}
\includegraphics[width=1.0\textwidth]{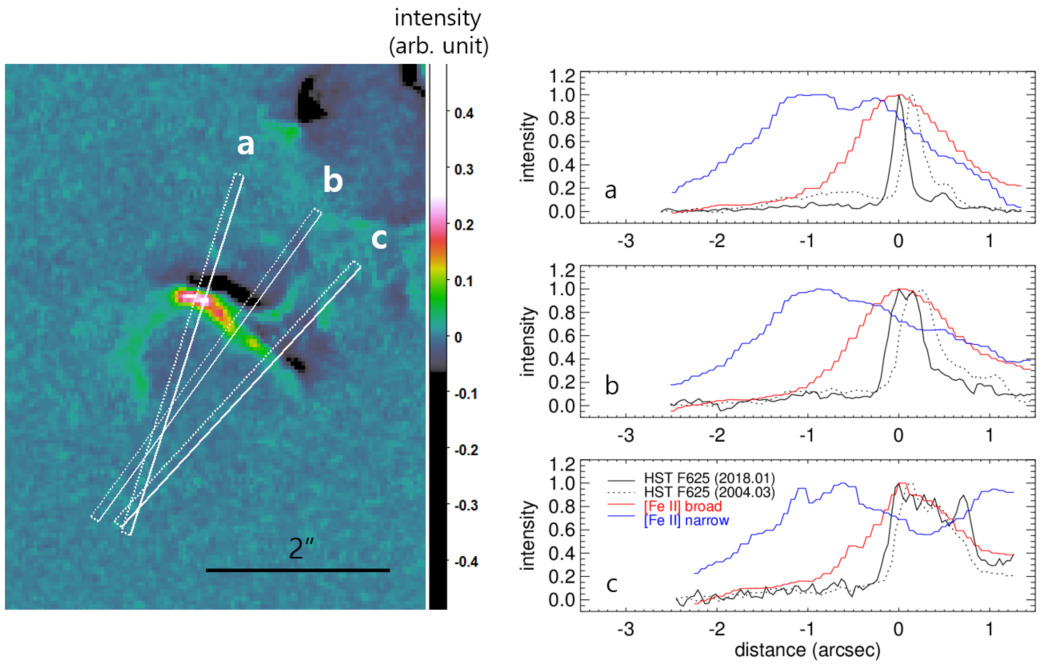} 
\caption{\linespread{1.0}\selectfont{}
{\bf Proper motion of the NW shock.} 
(left) Difference image produced by subtracting 
the 2004 {\it HST} ACS/WFC F625W image$^{1}$ from the 2018 {\it HST} WFC3/UVIS F625W image 
(Program ID 15337, P.I. Robert Fesen).
The scale bar corresponds to $2''$ (= 0.033 pc at 3.4 kpc). 
The uncertainty in this relative astrometric calibration is about a half pixel ($0\farcs03$).
(right) One-dimensional intensity profiles along the 
`slits' a, b, and c in the left panel.
The intensities have been normalized to their maximum values. 
The distance is measured from the maximum brightness position in the 2018 {\it HST} image (positive in NW).
The positional shift of the intensity peak positions is in the range      
$0\farcs09$ (slit c) to $0\farcs13$ (slits a and b), which corresponds to a proper motion of 
$0.006$--0.009 arcsecs yr$^{-1}$ (or 100--150~\kms).
}
\end{center}
\end{figure}

\clearpage
\newpage

\begin{figure}
\begin{center}
\includegraphics[width=0.5\textwidth]{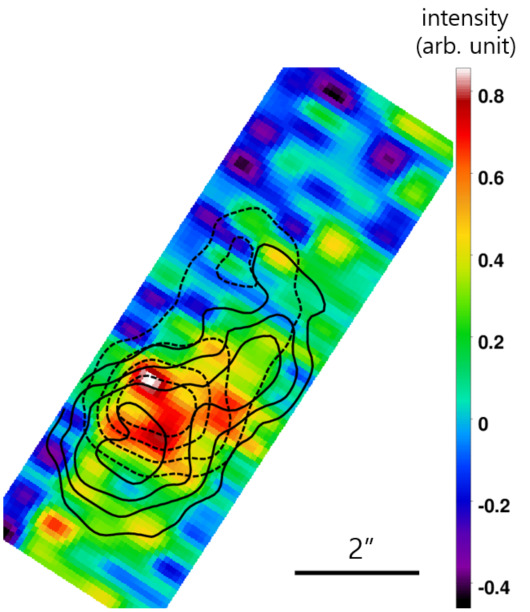} 
\caption{\linespread{1.0}\selectfont{}
{\bf Integrated intensity Br$\gamma$ map of Knot 24.}
The integrated velocity range is  
$\pm 15$~\kms\ with respect to the velocity 
center of the NLC of the Br$\gamma$ line ($-52$~\kms).
Since the BLC is much fainter than the NLC in this 
velocity range (see Fig. 2c), this map 
basically shows the distribution of the NLC. 
The intensity scale is linear in arbitrary units. The solid and dotted contours represent the 
Clump A-NLC and -BLC in \fetwoline\ emission (see Fig. 3).  
The scale bar corresponds to $2''$ (= 0.033 pc at 3.4 kpc). }
\end{center}
\end{figure}

\clearpage
\newpage

\begin{figure}
\begin{center}
\includegraphics[width=0.6\textwidth]{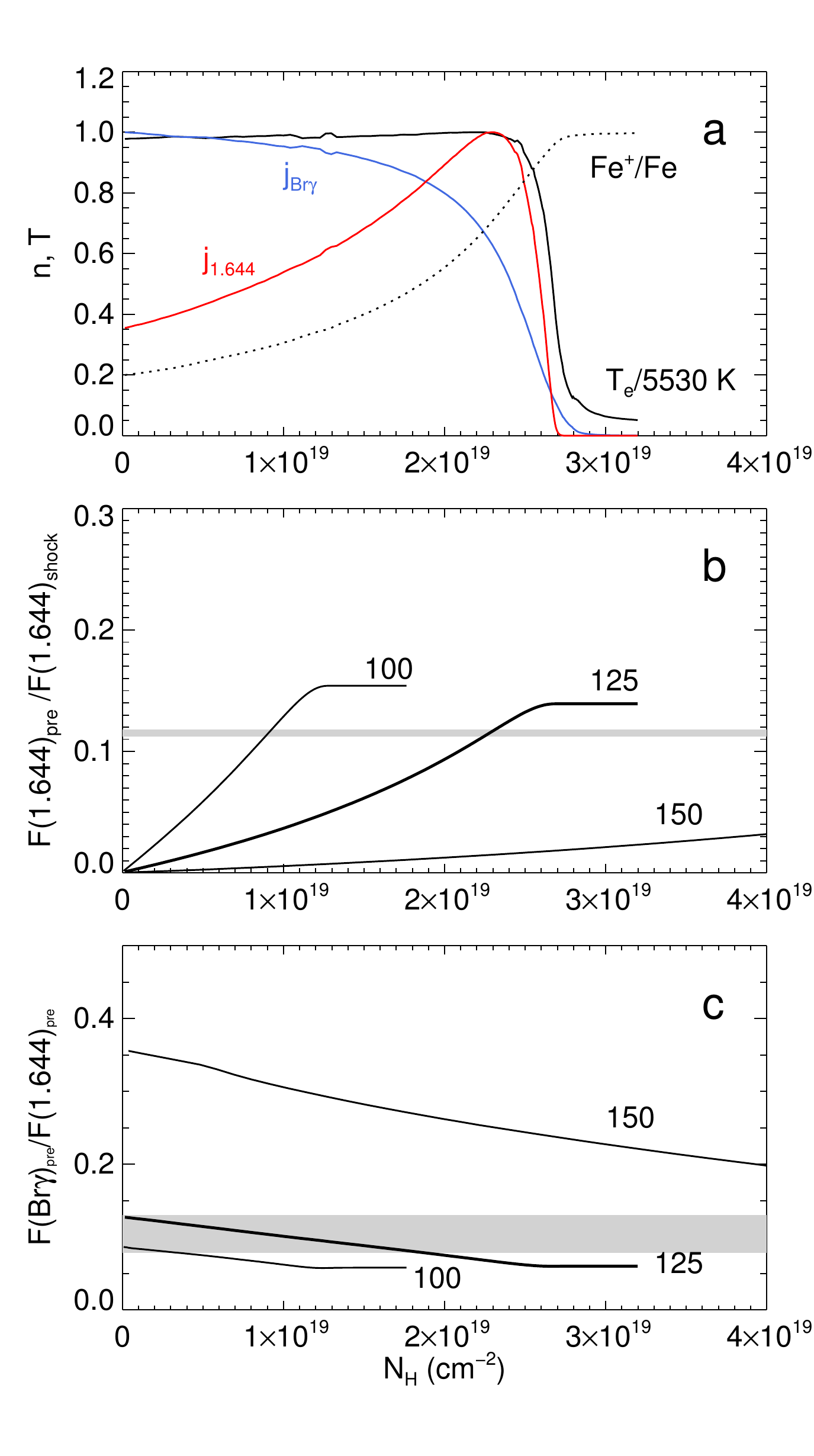} % this command will be ignored
\vspace*{-5mm}
\caption{\linespread{1.0}\selectfont{}
{\bf Shock precursor emission model.} 
{\bf a}, Physical and emission structures of radiative precursor produced by a 125~\kms\ plane-parallel shock. 
The ambient medium has H density $n_0=100$~cm$^{-3}$ and solar abundance$^{2}$.
The abscissa is H-nuclei column density from the shock front ($\nhcol$). 
The profiles of temperature and the fraction of Fe in Fe$^+$ (Fe$^+$/Fe) 
from MAPPINGS III Library$^{3}$ are shown.
Also shown are the normalized emissivities of \fetwoline\ and Br$\gamma$ lines calculated from the physical structure.
{\bf b}, \fetwoline\ flux of precursors of 100, 125, and 150~\kms\ shocks truncated at 
$\nhcol$ (Equation 3). It is divided by the \fetwoline\ flux from the shock. 
The shaded area represents the observed $\fluxnarrow/\fluxbroad$ ratio of Clump A. 
{\bf c},  Flux ratio of \brg\ to \fetwoline\ lines of precursors truncated 
at $\nhcol$.
The shaded area represents the observed $\fluxbrgnarrow/\fluxnarrow$ ratio of Clump A. 
}
\end{center}
\end{figure}

\clearpage

\noindent{\bf\large References}
\begin{enumerate}[1.]
\linespread{1.2}\selectfont{}
\item Fesen, R. A. {\it et al.} Discovery of outlying high-velocity oxygen-rich ejecta in Cassiopeia A. {\it Astrophys. J.} {\bf 636,} 859-872 (2006).
\item Asplund, M., Grevesse, N., Sauval, A. J. \& Scott, P. The chemical composition of the Sun. {\it Annu. Rev. Astron. Astr.} {\bf 47,} 481-522 (2009).
\item Allen, M. G., Groves, B. A., Dopita, M. A., Sutherland, R. S. \& Kewley, L. J. The MAPPINGS III library of fast radiative shock models. {\it Astrophys. J. Suppl. S.} {\bf 178,} 20-55 (2008).
\end{enumerate}

\end{document}